\documentclass[12pt, draftclsnofoot, onecolumn]{IEEEtran}
\usepackage{amsfonts,amsmath,amssymb}
\usepackage{graphicx,color,epsfig,rotating}
\usepackage{latexsym}
\usepackage{subfigure}
\usepackage{cite}
\usepackage{epic,eepic}
\usepackage{algorithm}
\usepackage{algorithmic}
\usepackage{float}
\usepackage{multirow}
\usepackage{setspace}
\newtheorem{theorem}{Theorem}

\newfont{\bb}{msbm10 scaled 1100}

\newcommand{\be}{\begin{equation}}
\newcommand{\ee}{\end{equation}}
\newcommand{\bea}{\begin{eqnarray}}
\newcommand{\eea}{\end{eqnarray}}
\newcommand{\bitem}{\begin{itemize}}
\newcommand{\eitem}{\end{itemize}}
\newcommand{\benum}{\begin{enumerate}}
\newcommand{\eenum}{\end{enumerate}}

\begin{document}
\title{Saturation Power based Simple Energy Efficiency Maximization Schemes for MU-MISO Systems}

\author{\IEEEauthorblockN{Jaehoon Jung, Sang-Rim Lee, and Inkyu Lee, \textit{Senior Member, IEEE}} \\
\IEEEauthorblockA{School of Electrical Eng., Korea University, Seoul, Korea\\
   Email: \{jhnjung, sangrim78, inkyu\}@korea.ac.kr} \\
}
\maketitle

\begin{abstract}
In this paper, we investigate an energy efficiency (EE) maximization problem in multi-user multiple input single output downlink channels.
The optimization problem in this system model is difficult to solve in general, since it is in non-convex fractional form.
Hence, conventional algorithms have addressed the problem in an iterative manner for each channel realization, which leads to high computational complexity.
To tackle this complexity issue, we propose a new simple method by utilizing the fact that the EE maximization is identical to the spectral efficiency (SE) maximization for the region of the power below a certain transmit power referred to as \textit{saturation power}.
In order to calculate the saturation power, we first introduce upper and lower bounds of the EE performance by adopting a maximal ratio transmission beamforming strategy.
Then, we propose an efficient way to compute the saturation power for the EE maximization problem.
Once we determine the saturation power corresponding to the maximum EE in advance, we can solve the EE maximization problem with SE maximization schemes with low complexity.
The derived saturation power is parameterized by employing random matrix theory, which relies only on the second order channel statistics.
Hence, this approach requires much lower computational complexity compared to a conventional scheme which exploits instantaneous channel state information, and provides insight on the saturation power.
Numerical results validate that the proposed algorithm achieves near optimal EE performance with significantly reduced complexity.
\end{abstract}


\section{Introduction} \label{intro}
\renewcommand\thefootnote{\fnsymbol{footnote}}
\footnotetext{The material in this paper was presented in part at the
IEEE International Conference on Communications (ICC), London, UK, June 2015 \cite{JJH:15}.}
\renewcommand\thefootnote{\arabic{footnote}}

Exponentially increasing service demands for wireless communications have mainly required huch higher transmission rate, which leads to increased energy consumption \cite{SHPark:12.1,HWPark:13}.
Recently, the energy consumption has been regarded as a crucial parameter when designing wireless networks, since low energy efficient transmission has a negative impact on the environment and hamper sustainable development.
Thus, from the perspective of green communications, energy efficiency (EE) has received a lot of attentions for future wireless communication systems \cite{Chen:11}.
The EE is defined as the ratio of the sum rate to the total power consumption measured in bit/Joule.

Many researches have addressed EE solutions for various system model scenarios \cite{Isheden:12,Kwan:12,Kwan:12_1,Heliot:12,HJKim:13,He:13,Bjornson:13,Ngo:13,SRLee:13}.
In \cite{Xu:13}, the EE problem was formulated by exploiting dirty paper coding and the uplink-downlink duality in broadcasting channels.
While this work presented a performance upper bound for broadcasting channels, many practical constraints exist due to high complexity.
For general scenarios with inter-user interference (IUI), the optimization problem for EE remains non-convex, and thus it is difficult and more challenging to solve.
Recently, EE schemes based on linear beamforming were studied for multiple-input single-output (MISO) interfering broadcasting channels \cite{He:13}.
By transforming the fractional programming into linear programming \cite{Isheden:12} and applying the weighted minimum mean square error (WMMSE) approach in \cite{Shi:11}, a local optimal solution was obtained in \cite{He:13}.
However, this algorithm solved the EE problem in an iterative manner for each channel realization that gives rise to high computational complexity.
Moreover, it is difficult to get insight on the system performance without resorting to Monte Carlo simulations.

To tackle these issues mentioned above, we investigate a simple and practical EE maximization scheme in multi-user (MU) MISO downlink channels.
First, we observe that the EE value obtained from the EE maximization problem is saturated at a certain transmit power, which will be referred to as \textit{saturation power}.
Then, the problem of the EE maximization becomes identical to that of the spectral efficiency (SE) maximization for the region below.
As a result, the EE maximization problem can simply be computed from the SE maximization by identifying the saturation power.

However, the optimum saturation power for the EE maximization scheme in considered system models is difficult to compute.
Hence, we first attempt to derive lower and upper bounds of the EE performance by applying maximal ratio transmission (MRT) beamforming.
Then, the saturation power of the lower and upper bounds of the EE are presented in closed form by employing random matrix theory \cite{Couillet:09,Huh:11,Hoydis:13,Zakhour:13}.
Here, based on the relationship of the derived saturation power and the EE performance, we can efficiently determine the saturation power by exploiting an interpolation method.
It is noted that the optimal saturation power is bounded by the derived saturation power for the lower and upper bounds of the EE.
Consequently, utilizing the derived saturation power, we can solve the EE problem efficiently by only adopting the SE maximization scheme.
Numerical results validate that the proposed algorithm achieves near optimal EE performance with much lower complexity.

The rest of the paper is comprised as follows: Section II presents a system model and the problem formulation.
In Section III, the relationship between EE and SE is described briefly.
Then, we derive the saturation power based on large system analysis and suggest a simplified scheme for the EE maximization utilizing the derived saturation power in Section IV.
From the simulation results in Section V, we confirm the validity of the proposed method. Finally, this paper is terminated with conclusions in Section VI.

Throughout the paper, we adopt lowercase and uppercase boldface letters for vectors and matrices, respectively. The superscript $(\cdot)^H$ stands for conjugate transpose. In addition, $||\cdot||$ and tr$(\cdot)$ represent Euclidean 2-norm and trace, respectively.
Also, $\bold{I}_d$ denotes an identity matrix of size $d$. A set of $N$ dimensional complex column vectors is expressed by $\mathbb{C}^N$.

\section{System Model} \label{sec:system}
In this paper, we consider an MU-MISO channel with bandwidth $W$ where a base station (BS) equipped with $M$ transmit antennas serves $N$ users with a single antenna.
Then, the received signal $y_k$ at the $k$-th user ($k = 1, \cdots, N$) is given by
\bea
y_k = \sqrt{p_k}\bold{h}_k^H \bold{v}_k s_k + \sum_{j \neq k}\sqrt{p_j}\bold{h}_k^H \bold{v}_j s_j + n_k\nonumber
\eea
where $p_k$ is the transmit power consumed by the $k$-th user satisfying $\sum_{k=1}^N p_k \leq P$ [Watt/Hz] in order to satisfy BS transmit power constraint $PW$, $\bold{h}_k \in \mathbb{C}^{M}$ defines the flat fading channel vector from the BS to the $k$-th user with the coherence time $T$, $\bold{v}_k$ means the beamforming vector for the $k$-th user with $||\bold{v}_k||^2 = 1$, $s_k \sim \mathcal{CN}(0,1)$ represents the complex data symbol intended for the $k$-th user, and $n_k \sim \mathcal{CN}(0,\sigma^2)$ stands for the additive white Gaussian noise at the $k$-th user.

For notational conveniences, we denote $\{\bold{p}\}$ and $\{\bold{v}\}$ as a set of all transmit power values and beamforming vectors, respectively.
Then, the individual rate of the $k$-th user is computed as
\bea
R_k(\{\bold{p}\}, \{\bold{v}\}) = \text{log}(1 + \text{SINR}_k(\{\bold{p}\}, \{\bold{v}\}))\nonumber
\eea
where $\text{SINR}_k(\{\bold{p}\}, \{\bold{v}\})$ indicates the individual signal-to-interference-plus-noise-ratio (SINR) for the $k$-th user as
\bea
\text{SINR}_k(\{\bold{p}\}, \{\bold{v}\}) = \frac{|\bold{h}_k^H \bold{v}_k|^2 p_k}{\sum_{j \neq k}|\bold{h}_k^H \bold{v}_j|^2 p_j + \mathcal{N}_0}.\nonumber
\eea
Here, $\mathcal{N}_0$ represents $\mathcal{N}_0 = \sigma^2/W$.
During a time-frequency block $TW$, the total amount of the transmitted information is given by
\bea
TW\sum_{k}\log_2(1 + \text{SINR}_k).~~~[\text{bits}]\nonumber
\eea

From an EE point of view, we consider the power consumption for a BS \cite{He:13, Bjornson:15}, where the total power consumption during the time-frequency block $TW$ is modeled as
\bea
P_T(\{\bold{p}\}) = TW\left(\xi\sum_k p_k ||\bold{v}_k||^2 + P_{const}\right).~~~[\text{Joule}]\label{P_T}
\eea
Here, $\xi \geq 1$ stands for an inefficiency of the power amplifier and $P_{const}$ equals $P_{const} = MP_c + P_o$ where $P_c$ is defined as $P_c = \frac{P_c^{'}}{W}$ with $P_c^{'}$ being the constant circuit power consumption proportional to the number of radio frequency chains, and $P_o$ means $P_o = \frac{P_o^{'}}{W}$ with $P_o^{'}$ indicating the static power at the BS which is independent of the number of transmit antennas.

Then, the EE is defined as the ratio of the sum rate to the total power consumption
\bea
\text{EE}(\{\bold{p}\}, \{\bold{v}\}) = \frac{\sum_k R_k(\{\bold{p}\}, \{\bold{v}\})}{P_T(\{\bold{p}\})}.\nonumber
\eea
Therefore, the EE maximization problem can be formulated by
\bea
\max_{\{\bold{p}\}, \{\bold{v}\}}&& \text{EE}(\{\bold{p}\}, \{\bold{v}\}) \nonumber\\
\text{s.t.}~&& \sum_{k=1}^N p_k \leq P.\label{EE}
\eea

It is noted that problem (\ref{EE}) is non-convex because of coupled interference and the fractional form, and thus computing a solution of the problem is quite complicated.
In \cite{He:13}, a local optimal solution of the EE for interfering broadcasting channels was obtained by two layer optimization adopting a linear subtractive form.
However, it should be solved in an iterative manner for each channel realization, which gives rise to high computational complexity.
In what follows, we focus on a simple algorithm which can solve the EE maximization with reduced complexity.

\section{Properties of Energy Efficiency} \label{Capacity Analysis}

In this section, we investigate the characteristics of the EE.
Based on the properties of the EE described in this section, the derivation of the saturation power is triggered to optimize the EE performance in a simple manner.
It is interesting to note that the EE performance is saturated once the total transmit power exceeds a certain point, which we call saturation power.
Then, the maximization of the EE becomes identical to that of the SE for the region below the saturation power.
To explain this, we consider a simple EE model for the transmit power $P$ as
\bea
\text{EE}(P) = \frac{R(P)}{P+P_{static}}\label{simple EE}
\eea
where $R(P) = \text{log}(1+P)$ and $P_{static}$ indicates the static power consumption term.

From this EE expression (\ref{simple EE}), the optimal transmit power $P_{EE}$ which maximizes the EE can be calculated in closed form as \cite{Isheden:12}
\bea
P_{EE} = \text{exp}\left(\mathcal{W}_0\left(\frac{P_{static}-1}{e}\right)+1\right)-1\nonumber
\eea
where $\mathcal{W}_0(\cdot)$ denotes the principal branch of the Lambert W function defined as the inverse function of $f(x) = xe^x$.

For the transmit power region below $P_{EE}$, full transmit power should be applied to achieve the maximal performance of the EE.
This is due to the fact that when the total transmit power is fully consumed at the region below the saturation power, the consumed power becomes constant which does not affect the EE optimization.
In this case, the considered problem is equivalent to the sum rate maximization.
This suggests that transmitting the maximum available power is most energy efficient at this region.
In the same way, the SE performance can be maximized at the same region because the rate $R(P)$ is monotonically increasing function with respect to $P$.
In contrast, for the region above the saturation power, consuming full power at the BS degrades the EE performance, since a sum rate gain cannot compensate for the increased power consumption in the EE.

In Figure 1, we illustrate the performance curves of the SE and EE for equation (\ref{simple EE}) with respect to the transmit power $P$.
In this example, the saturation power $P_{EE}$ is shown to be about 2 dB.
It can be observed that the SE maximization is identical to the EE maximization when the transmit power $P$ is smaller than the saturation power $P_{EE}$.
Also, $P_{EE}$ corresponds to the power which yields the maximal SE.
In Figure 1 (a), the EE scheme achieves the same rate as the SE scheme for $P \leq P_{EE}$, while the rate of the EE scheme becomes saturated for $P \geq P_{EE}$, since the EE scheme fixes the power at $P_{EE}$ to maximize the EE.
Meanwhile, the SE algorithm always transmits at full power for maximizing the SE even after the saturation power $P_{EE}$.
In Figure 1 (b), the SE scheme exhibits a performance loss in terms of of the EE because a gain of the rate cannot make up for the impact of the increased power consumption as mentioned before.
\begin{figure}[t]\centering
\includegraphics[width=5in]{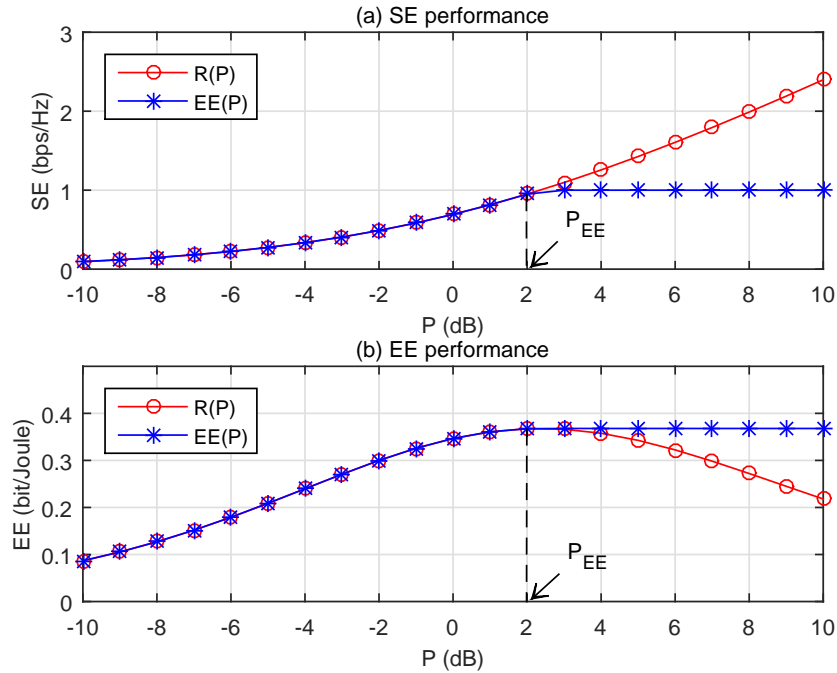}
\caption{Comparison of SE and EE for equation (\ref{simple EE}) in terms of the transmit power $P$ (a) SE performance and (b) EE performance}
\end{figure}

\section{Derivation of Saturation Power}
In this section, motivated by properties of the EE shown in Section III, we first focus on determining the saturation power which starts to yield the saturated EE performance.
Unfortunately, the EE solution in \cite{He:13} requires iterative methods, and thus it is not possible to obtain the saturation power in closed form.
Alternatively, we address lower and upper bounds of the EE performance to allow simple computations of the saturation power.
From the derived lower and upper bounds of the EE, we can identify the saturation power corresponding to each bound of the EE in a closed form solution.
Then, by utilizing the determined saturation power, we propose a new EE maximization scheme only adopting the SE maximization method with reduced complexity.
For mathematical tractability, we assume that the transmit power of each user $p_k$ for $k = 1, \cdots, N$ is equally allocated.

\subsection{Saturation Power for a Lower Bound of EE}

First, we start with obtaining a lower bound of the EE performance.
One simple precoding which can serve as a lower bound is MRT beamforming which employs $\bold{v}_{k,MRT} = \frac{\bold{h}_k}{||\bold{h}_k||}$.
In this case, the EE for MRT $\eta_{MRT}$ with equal power allocation can be expressed as
\bea
\eta_{MRT} = \frac{\sum_{k=1}^N \log(1+\text{SINR}_{k,MRT})}{\xi P + P_{const}}\label{eta_MRT}
\eea
where $\text{SINR}_{k,MRT}$ is given by
\bea
\text{SINR}_{k,MRT} = \frac{|\bold{h}_k^H\bold{v}_{k,MRT}|^2\frac{P}{N}}{\sum_{j \neq k}|\bold{h}_k^H\bold{v}_{j,MRT}|^2\frac{P}{N} + \mathcal{N}_0 }.\label{SINR_MRT}
\eea

It is clear that $\text{SINR}_{k,MRT}$ changes for every channel realizations.
To avoid calculating these instantaneous channel gains, we apply random matrix theory in (\ref{SINR_MRT}).
It is worth noting that in the asymptotic regime, the channel gain values become deterministic which depends only on the second order channel statistics and the randomness according to instantaneous channel realizations disappears.
Although the parameters are obtained in the large system limit, we will show in the simulation section that this approximation is well matched even for small dimensions.

To derive large system results, we will utilize the lemma in \cite{Wagner:12} which assumes that the number of users $N$ and the number of transmit antennas $M$ grow large with $\frac{N}{M}$ at a fixed ratio.
We emphasize that the asymptotic analysis is used only to derive deterministic channel gain values, and the system which we consider in this paper has a finite dimension.
From \cite{Wagner:12}, we can calculate the deterministic value for both the desired signal power $|\bold{h}_k^H\bold{v}_{k,MRT}|^2$ and the interference signal power $|\bold{h}_k^H\bold{v}_{j,MRT}|^2$ in (\ref{SINR_MRT}).
First, the desired signal power for MRT is written by
\bea
|\bold{h}_k^H\bold{v}_{k,MRT}|^2=\frac{|\bold{h}_k^H \bold{I}_M \bold{h}_k|^2}{||\bold{h}_k||^2} = \bold{h}_k^H \bold{I}_M \bold{h}_k.\nonumber
\eea
Then, using the trace lemma in \cite{Debbah:11}, we have
\bea
\bold{h}_k^H \bold{I}_M \bold{h}_k - \text{tr}(\bold{I}_M) \xrightarrow{a.s.} 0.\label{Desired}
\eea

Also, the interference signal power for MRT is determined in a similar manner as
\bea
|\bold{h}_k^H\bold{v}_{j,MRT}|^2=\frac{|\bold{h}_k^H \bold{I}_M \bold{h}_j|^2}{||\bold{h}_j||^2} = \frac{\bold{h}_k^H \bold{I}_M \bold{h}_j\bold{h}_j^H \bold{I}_M\bold{h}_k}{\bold{h}_j^H \bold{h}_j}.\label{eq:if_sig}
\eea
After some mathematical manipulations, the numerator and denominator terms for the interference signal power in (\ref{eq:if_sig}) converge almost surely as
\bea
\bold{h}_k^H \bold{I}_M \bold{h}_j\bold{h}_j^H \bold{I}_M\bold{h}_k &-& \text{tr}({\bold{I}_M}^2) \xrightarrow{a.s.} 0,\nonumber\\
\bold{h}_j^H \bold{h}_j &-& \text{tr}({\bold{I}_M}) \xrightarrow{a.s.} 0.\nonumber
\eea
Then, the interference signal power $|\bold{h}_k^H\bold{v}_{j,MRT}|^2$ is given by
\bea
\frac{\bold{h}_k^H \bold{I}_M \bold{h}_j\bold{h}_j^H \bold{I}_M\bold{h}_j}{\bold{h}_j^H \bold{h}_j} - 1 \xrightarrow{a.s.} 0.\label{Interference}
\eea

By replacing the deterministic channel gain values in (\ref{Desired}) and (\ref{Interference}) into $\text{SINR}_{k,MRT}$, the asymptotic EE for MRT $\eta_{MRT}^\circ$ is presented as
\bea
\eta_{MRT}^\circ = \frac{\sum_{k=1}^N \log(1+\text{SINR}_{k,MRT}^\circ)}{\xi P + P_{const}}\label{asymptotic_EE_MRT}
\eea
where $\text{SINR}_{k,MRT}^\circ$ is denoted as
\bea
\text{SINR}_{k,MRT}^\circ = \frac{M P}{(N-1)P + N\mathcal{N}_0}.\nonumber
\eea
We can note that $\text{SINR}_{k,MRT}^\circ$ is a function of $P$ and is no longer dependent on channel realizations.
As a result, $\eta_{MRT}^\circ$ can be identified only based on given system configurations.
%
To obtain the saturation power for a lower bound of the EE, we utilize equation (\ref{asymptotic_EE_MRT}) for the following theorem.
\begin{theorem}
A lower bound of $\eta_{MRT}^\circ$, defined by $\eta_{LB}$, is expressed by
\bea
\eta_{LB} = \frac{N M P}{(\xi P + P_{const})\{(N + M -1) P + N\mathcal{N}_0\}}.\nonumber
\eea
Then, the saturation power $P_{LB}$ corresponding to the EE lower bound $\eta_{LB}$ is computed by
\bea
P_{LB} = \sqrt{\frac{N \mathcal{N}_0 P_{const}}{\xi(N + M -1)}}.\label{P_LB}
\eea
\end{theorem}
\begin{IEEEproof}
The numerator term of (\ref{asymptotic_EE_MRT}) is reformulated by
\bea
N \log\left(\!\!1\!+\! \frac{M P}{(N-1) P\!+\! N\mathcal{N}_0}\!\!\right)\!\!\!\!\!&=&\!\!\!\!N \log \frac{(N+M-1) P\!+\! N\mathcal{N}_0}{(N-1) P \!+\! N\mathcal{N}_0}\nonumber\\
&=&\!\!\!\!\!-N \log\left(\!\!1 \!-\! \frac{M P}{(N+M-1) P \!+\! N\mathcal{N}_0}\!\!\right)\!\!.\label{log}
\eea
From the fact that the term $\frac{M P}{(N+M-1) P + N\mathcal{N}_0}$ is smaller than 1, equation (\ref{log}) can be bounded by adopting the relationship $\log(1+x) \leq x $ for $|x| < 1$ as
\bea
-N \log\left(1 - \frac{M P}{(N+M-1) P + N\mathcal{N}_0}\right) \geq \frac{N M P}{(N+M-1) P + N\mathcal{N}_0} \triangleq R_{LB}.\label{R_LB}
\eea

Consequently, the lower bounded EE $\eta_{LB}$ is presented by
\bea
\eta_{LB} = \frac{R_{LB}}{\xi P + P_{const}}.\label{eq:EE_LB}
\eea
Then, the saturation power $P_{LB}$ can be determined by differentiating $\eta_{LB}$ in (\ref{eq:EE_LB}) with respect to $P$.
Thus, $P_{LB}$ which maximizes $\eta_{LB}$ is calculated from the following equation as
\bea
(N+M-1)\xi P_{LB}^2 = N\mathcal{N}_0 P_{const}.\nonumber
\eea
From this equation, we arrive at (\ref{P_LB})
\end{IEEEproof}
According to the result in Theorem 1, the saturation power for a lower bound of EE is obtained with a closed form expression (\ref{P_LB}).

\subsection{Saturation Power for an Upper Bound of EE}

In the previous subsection, a lower bound for the EE is found by applying MRT with equal power allocation.
Now, we derive an upper bound of the EE by ignoring the effect of IUI.
Then, the EE with no IUI, $\eta_{no-IUI}$, can be given by
\bea
\eta_{no-IUI} = \frac{\sum_{k=1}^N \log\left(1+\frac{|\bold{h}_k^H\bold{v}_{k,MRT}|^2}{\mathcal{N}_0}\frac{P}{N}\right)}{\xi P + P_{const}}.\label{eta_max}
\eea
When the IUI is not considered, the numerator term of $\eta_{no-IUI}$ is maximized by the MRT beamforming because the beam is aligned with the channel for the intended user.
Therefore, the EE performance is upper bounded by $\eta_{no-IUI}$.

By employing the large system analysis as in Section IV-A, the numerator of $\eta_{no-IUI}$ in (\ref{eta_max}) is presented as
\bea
\log\left(1+\frac{|\bold{h}_k^H\bold{v}_{k,MRT}|^2}{\mathcal{N}_0}\frac{P}{N}\right) - R_{UB}\xrightarrow{a.s.} 0\label{R_UB}
\eea
where $R_{UB}$ is defined as $R_{UB} \triangleq N \log\left(1+\frac{M}{N \mathcal{N}_0}P\right)$.
Then, an asymptotic upper bound of the EE, denoted by $\eta_{UB}$, is expressed as
\bea
\eta_{UB} = \frac{R_{UB}}{\xi P + P_{const}}.\label{eta_UB}
\eea
To compute the saturation power for the EE upper bound $\eta_{UB}$, we address the following theorem.
\begin{theorem}
The saturation power $P_{UB}$ which maximizes $\eta_{UB}$ is written by
\bea
P_{UB} = \frac{N \mathcal{N}_0}{M}\left[\exp\left(1 + \mathcal{W}_0\left(\frac{1}{e}\left(\frac{M P_{const}}{N \mathcal{N}_0 \xi}-1\!\right)\!\right)\!\right)\!-1\right].\label{P_UB}
\eea
\end{theorem}
\begin{IEEEproof}
By differentiating $\eta_{UB}$ with respect to the total transmit power $P$, it follows
\bea
\frac{d \eta_{UB}}{d P} = \frac{N}{(\zeta P + P_{const})^2} \left[ \frac{M}{N\mathcal{N}_0}\frac{\xi P + P_{const}}{1+\frac{M}{N\mathcal{N}_0}P} - \xi \log\left(1+\frac{M}{N\mathcal{N}_0}P\right) \right].\label{eq:EE_UB_diff}
\eea
For the saturation power $P_{UB}$ corresponding to $\eta_{UB}$, setting equation (\ref{eq:EE_UB_diff}) to zero yields
\bea
\frac{M}{N \mathcal{N}_0}(\xi P_{UB} + P_{const}) - \xi s\log s = 0\nonumber
\eea
where $s =1+\frac{M}{N \mathcal{N}_0}P_{UB}$.

Then, we have
\bea
\frac{M P_{const}}{N \mathcal{N}_0 \xi} - 1 = s(\log s - 1).\nonumber
\eea
With the Lambert W function, this form can be solved by a closed form expression as
\bea
\log s = 1 + \mathcal{W}_0\left(\frac{1}{e}\left(\frac{M P_{const}}{N \mathcal{N}_0 \xi}-1\right)\right).\nonumber
\eea
Since $\mathcal{W}\left(\frac{1}{e}\left(\frac{M P_{const}}{N \mathcal{N}_0 \xi}-1\right)\right) \geq -1$, which is guaranteed by $s \geq 1$ for $P_{UB} \geq 0$, the principal branch of the Lambert W function $\mathcal{W}_0$ is selected.
From this equation, we can reach the saturation power $P_{UB}$ in (\ref{P_UB}).
\end{IEEEproof}

\subsection{Relationship among $P_{LB}$, $P_{UB}$, and the optimal saturation power $P^*$}

From the previous subsections, the saturation power for lower and upper bounds of the EE has been derived.
In this subsection, we address the property of the optimal saturation power denoted as $P^*$ related to $P_{LB}$ and $P_{UB}$.
As already mentioned, the average sum rate of the SE maximization scheme denoted by $R_{SE}$ is clearly bounded between $R_{LB}$ in (\ref{R_LB}) and $R_{UB}$ in (\ref{R_UB}).
From this, we will show that the optimal saturation power $P^*$ lies between $P_{LB}$ and $P_{UB}$ by adopting the linear parametric programming approach.

It can be seen that the optimization problem (\ref{EE}) belongs to fractional programming.
Hence, this problem can be transformed into parametric programming as in \cite{Isheden:12}.
We consider the following equivalent form of the fractional program in (\ref{EE}) as
\bea
\max_{\{\bold{p}\},~ \{\bold{v}\},~ \lambda \in \mathbb{R}}&& \lambda \nonumber\\
\text{s.t.}~~~~~&& \sum_k R_k(\{\bold{p}\}, \{\bold{v}\}) - \lambda P_T(\{\bold{p}\}) \geq 0.\nonumber
\eea
For a given parameter $\lambda$, it is noted that the optimization problem is referred to as a feasibility problem in $\{\bold{p}\}$ and $\{\bold{v}\}$.
Therefore, the optimal value of the parameter $\lambda$ can be found by using a bisection method for the feasibility problem at each step of the algorithm \cite{Boyd:04}.

Defining a function $F(\lambda)$ as
\bea
F(\lambda) = \max_{\{\bold{p}\}, \{\bold{v}\}} \sum_k R_k(\{\bold{p}\},\{\bold{v}\}) - \lambda P_T(\{\bold{p}\}),\nonumber
\eea
it is obvious that $F(\lambda)$ is convex and strictly decreasing in $\lambda$.
Moreover, this is regarded as bi-criterion optimization such that $\sum_k R_k(\{\bold{p}\},\{\bold{v}\})$ is maximized while $P_T(\{\bold{p}\})$ is minimized.
The parameter $\lambda$ determines the relative weight of the total power consumption $P_T(\{\bold{p}\})$.

For this bi-criterion problem, the set of Pareto-optimal values is called the optimal trade-off curve \cite{Isheden:12}.
As presented in \cite{He:13}, solving problem (\ref{EE}) is equivalent to finding the root of the nonlinear function $F(\lambda)$, i.e., $F(\lambda)=0$.
In other words, $\lambda$ means the slope of the tangent for the trade-off curve and the optimal $\lambda$ denoted by $\lambda^*$ occurs when $F(\lambda^*) = 0$.
\begin{figure}[t]\centering
\includegraphics[width=5in]{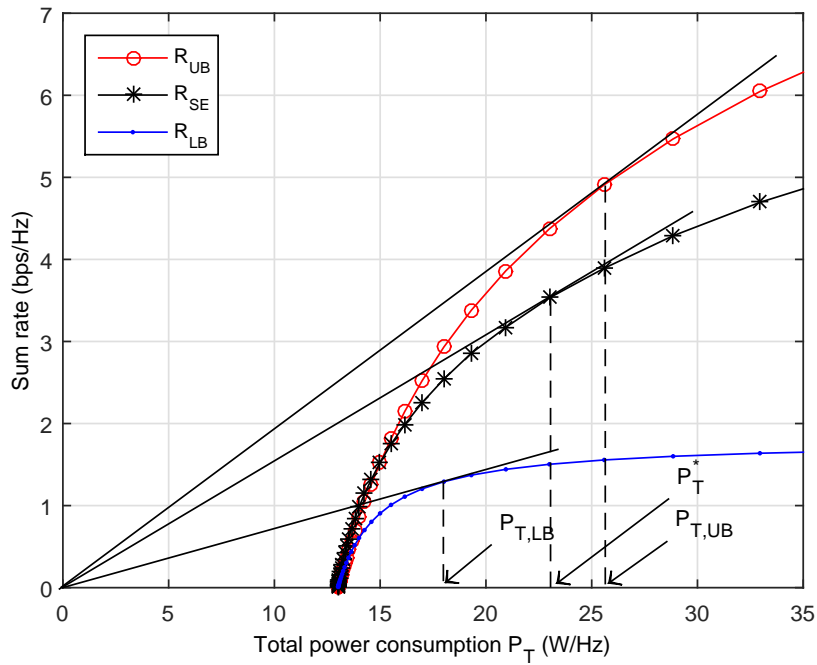}
\caption{Trade-off curves between sum rate and total consumed power with $P_c' = 30$ dBm and $P_o' = 40$ dBm}
\end{figure}
Then, by exploiting these properties of the linear parametric programming, we can address the relationship among $P_{LB}$, $P_{UB}$, and $P^*$.

From the trade-off curves in Figure 2, we can identify the slope of the tangent for $R_{LB}, R_{SE}$, and $R_{UB}$.
In this figure, the trade-off curve is illustrated between the sum rate and the total power consumption in (\ref{P_T}) with $N=M=3$ and $P_{const} = 13$.
We can see that the SE performance $R_{SE}$ is certainly bounded as $R_{LB} < R_{SE} < R_{UB}$ for all $P_T$.
Here, $P_{T,LB}$, $P_{T}^*$, and $P_{T,UB}$ denote the total power consumption at the BS for $R_{LB}$, $R_{SE}$, and $R_{UB}$, respectively.
It is noted that the optimal power consumption for each scheme is defined as a contact point with each sum rate curve and the corresponding tangent.

Then, the saturation power is calculated by subtracting $P_{const}$ from the optimal power consumption, e.g., $P_{LB} = P_{T,LB} - P_{const}$.
As shown in the figure, the saturation power for each scheme has the relationship of $P_{LB} < P^* < P_{UB}$.
Accordingly, the achieved EE values which equal the slope of the tangent for each sum rate are given as $\eta_{LB}(P_{LB}) < \eta_{SE}(P^*) < \eta_{UB}(P_{UB})$ where $\eta_{SE}(P^*) = \frac{R_{SE}(P^*)}{\zeta P^* + P_{const}}$ represents the maximum EE value for $\eta_{SE}$ corresponding to the saturation power $P^*$.
Therefore, utilizing this property, we quantify the optimal saturation power $P^*$ as a function of $P_{LB}$ and $P_{UB}$ in the following.

\subsection{Proposed EE Scheme based on the Saturation Power}

From the derived saturation power for lower and upper bounds of the EE, we will determine the saturation power $P^*$.
To this end, we adopt an interpolation method in a similar way to \cite{JHKim:09}.
First, the maximum value of $\eta_{SE}$ is defined by $\gamma_{SE} = \eta_{SE}(P^*)$.
Also, the maximum EE value for the lower and upper bounds $\gamma_{LB}$ and $\gamma_{UB}$ are denoted by $\eta_{LB}(P_{LB})$ and  $\eta_{UB}(P_{UB})$, respectively.
Then, the proposed saturation power $P_{prop}$ can be computed as a point between $P_{LB}$ and $P_{UB}$  by exploiting the relationship between the saturation power and the corresponding EE value for each bound.
For instance, if $\gamma_{SE}$ is close to $\gamma_{UB}$, the saturation power $P^*$ should be set near $P_{UB}$.

Therefore, we consider the following relation on the difference between various bounds addressed in Section IV-C as
\bea
\frac{\gamma_{UB}-\gamma_{SE}}{\gamma_{UB}-\gamma_{LB}}:\frac{\gamma_{SE}-\gamma_{LB}}{\gamma_{UB}-\gamma_{LB}}
=\frac{P_{UB}-P_{prop}}{P_{UB}-P_{LB}}:\frac{P_{prop}-P_{LB}}{P_{UB}-P_{LB}}.\label{ratio}
\eea
After some mathematical manipulations, equation (\ref{ratio}) is expressed as
\bea
P_{UB}-P_{prop} = G(P_{prop}-P_{LB}),\nonumber
\eea
where the EE gap $G$ represents
\bea
G = \frac{\gamma_{UB}-\gamma_{SE}}{\gamma_{SE}-\gamma_{LB}}.\label{EE_gap}
\eea
Then, the proposed saturation power $P_{prop}$ can be calculated as
\bea
P_{prop} = \omega P_{LB} + (1-\omega) P_{UB}\label{inter}
\eea
where $\omega = \frac{G}{1+G}$ means the weight factor between $P_{LB}$ and $P_{UB}$.

In fact, it is not possible to obtain $G$ directly since $\gamma_{SE}$ is unknown.
Therefore, in order to compute the EE gap $G$, we consider regularized zero forcing (RZF) beamforming \cite{Wagner:12}.
The relation between $\gamma_{SE}$ and the maximum EE performance for RZF $\gamma_{RZF}$ can be formulated by $\gamma_{SE} = \beta \gamma_{RZF}$, where a constant $\beta > 1$ accounts for the performance gain of $\eta_{SE}$ over $\eta_{RZF}$.
In what follows, we determine $\gamma_{RZF}$ which will be used in calculating $G$ in (\ref{EE_gap}) for a given $\beta$.

By adopting random matrix theory in \cite{Wagner:12}, the asymptotic EE performance for RZF $\eta_{RZF}^\circ$ is expressed as
\bea
\eta_{RZF}^\circ = \frac{\sum_{k=1}^N \log(1+\text{SINR}_{k,RZF}^\circ)}{\xi \sum_{k=1}^N p_k + P_{const}}\nonumber
\eea
where $\text{SINR}_{k,RZF}^\circ$ is obtained as
\bea
\text{SINR}_{k,RZF}^\circ = \frac{(m_k^\circ)^2}{\Gamma_k^\circ + \frac{\Psi^\circ}{\rho}(1+m_k^\circ)^2}.\nonumber
\eea
Here, the deterministic equivalent values $m_k^\circ$, $\Gamma_k^\circ$, and $\Psi^\circ$ are derived in \cite{Wagner:12} for RZF and $\rho = P /\mathcal{N}_0$.
These parameters are affected by $p_k$ for $\forall k$.

Assuming equal power allocation with uncorrelated channels, the deterministic equivalent for all users has the same value, i.e., $m_k^\circ = m^\circ$ and $\Gamma_k^\circ = \Gamma^\circ$.
Then, the $\eta_{RZF}^\circ$ with equal power allocation denoted by $\eta_{RZF}$ can be calculated as
\bea
\eta_{RZF} = \frac{N\log\left(1+\frac{(m^\circ)^2 P}{\Gamma^\circ P + \Psi^\circ(1+m^\circ)^2\mathcal{N}_0}\right)}{\xi P + P_{const}}.\nonumber
\eea
To determine the maximum EE for RZF $\gamma_{RZF}$, $\eta_{RZF}$ is differentiated with respect to $P$ and is set to zero as
\bea
\frac{d\eta_{RZF}}{d P} \!\!=\!\!\frac{N (m^\circ)^2 A}{\{((m^\circ)^2\!+\!\Gamma^\circ)P\!+\!A\}(\Gamma^\circ P\!+\!A)(\xi P\!+\!P_{const})}\!-\! N\! \log\!\left(\!\!1\!+\!\frac{(m^\circ)^2 P}{\Gamma^\circ P\!+\! A}\!\right)\!\!\frac{\xi}{(\xi P\!+\! P_{const})^2} \!=\! 0\label{diff_EE}
\eea
where $A = \Psi^\circ(1+m^\circ)^2 \mathcal{N}_0$.
Here, we define the function $f(P)$ to identify the saturation power $P_{RZF}$ for $\eta_{RZF}$ as
\bea
f(P) = \log\left(1 + \frac{(m^\circ)^2 P}{\Gamma^\circ P + A}\right) - \frac{(m^\circ)^2 A(P + \frac{P_{const}}{\xi})}{\{((m^\circ)^2+\Gamma^\circ)P + A\}(\Gamma^\circ P + A)}.\nonumber
\eea

It is interesting to note that the function $f(P)$ is monotonically increasing with respect to $P$ and the equation $f(P) = 0$ has a unique solution.
Also, $f(P)$ converges to $-\frac{(m^\circ)^2 P_{const}}{\zeta A}$ for $P \rightarrow 0$ and $\log(1 + (m^\circ)^2/\Gamma^\circ)$ for $P \rightarrow \infty$.
Therefore, the EE saturation power $P_{RZF}$ can be computed simply by one dimensional search and the maximum EE performance can be obtained as $\gamma_{RZF} = \eta_{RZF}(P_{RZF})$.
Consequently, after $\gamma_{SE}$ is replaced by $\beta \gamma_{RZF}$, we can determine the saturation power $P_{prop}$ in (\ref{inter}).

Now, we propose a simplified EE maximization scheme by utilizing the derived saturation power $P_{prop}$.
In the simplified scheme, a solution of the SE maximization problem is adopted for the original EE maximization problem.
First, when the available transmit power $P$ is less than $P_{prop}$, the SE maximization scheme is conducted with full power $P$ to maximize the EE performance.
On the contrary, if $P$ is greater than $P_{prop}$, the fixed transmit power $P_{prop}$ is used with the SE maximization method.
In summary, after the saturation power $P_{prop}$ is calculated in (\ref{inter}), the SE maximization algorithm in \cite{Shi:11} is processed with the transmit power given by $\min(P_{prop},P)$ to generate a beamforming solution.

Next, we briefly address the computational complexity.
The structure of the EE algorithm in \cite{He:13} is comprised by the outer layer and the inner layer optimization.
The outer layer searches for the EE parameter $\eta$, while the inner layer solves the non-fractional subtractive problem for a given $\eta$ computed at the outer layer.
Thus, the inner layer algorithm should be executed whenever the EE parameter is updated at the outer layer, and this causes high computational complexity.

The complexity of the algorithm in \cite{Shi:11} is similar to that of the inner layer part in \cite{He:13}.
It is noted that the SE maximization algorithm in \cite{Shi:11} is processed only once in the proposed EE scheme.
Moreover, when determining the saturation power $P_{prop}$, we have $P_{LB}$ and $P_{UB}$ in closed form which depends only on the second order channel statistics and the estimation of $\gamma_{SE}$ needs simple one-dimensional search.
Hence, the complexity of our proposed algorithm is much lower than that of the EE algorithm in \cite{He:13}.
The computation time of the algorithm in \cite{He:13} is contingent on the convergence threshold $\delta$ for the outer layer and it takes about ten times higher than that of the proposed scheme with $\delta = 10^{-3}$, while the EE performance of the proposed scheme is quite close to that of the algorithm in \cite{He:13}.

\section{Numerical Results} \label{sec:simulation}

\begin{figure}[t]\centering
\includegraphics[width=5in]{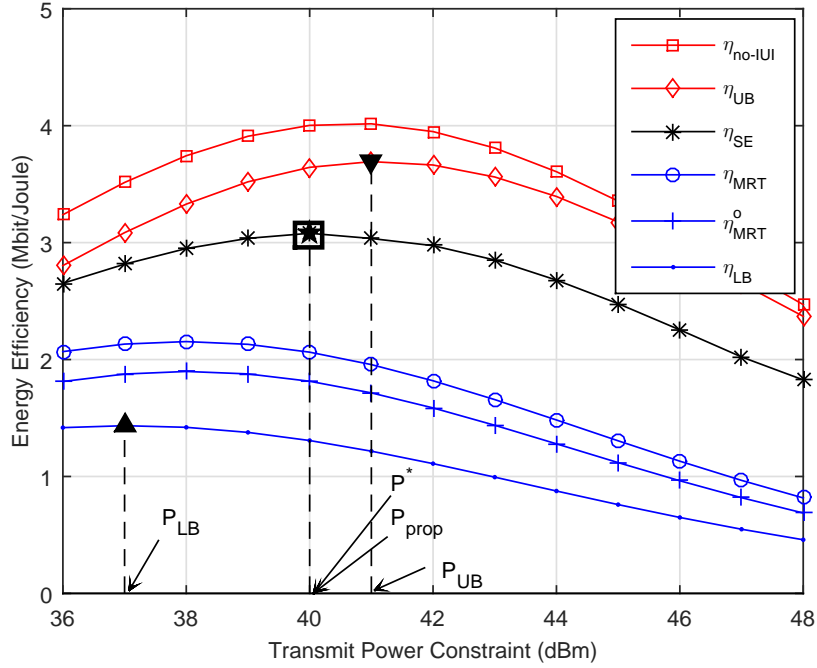}
\caption{Comparison of the saturation power with $P_c' = 30$ dBm and $P_o' = 40$ dBm}
\label{fig2}
\end{figure}
In this section, we verify the validity of our proposed method through Monte Carlo simulations.
The numbers of users $N$ and transmit antennas $M$ are equal to 3 unless specified otherwise.
Also, we adopt the bandwidth $W = 20$ MHz, the noise spectral density $\mathcal{N}_0 = -174$ dBm/Hz, noise figure $N_F = 7$ dB and the inefficiency of the power amplifier $\xi = 1$.
The circuit power per antenna $P_c'$ and the static power consumed at the BS $P_o'$ are set to 30 dBm and 40 dBm, respectively.

First, we evaluate the saturation power derived in Section IV in Figure 3.
Here, regular and inverted triangles mean the maximum EE for (\ref{P_LB}) and (\ref{P_UB}), respectively, and
$P_{LB}$ and $P_{UB}$ are computed as the derived saturation power corresponding to these maximum points shown in (\ref{P_LB}) and (\ref{P_UB}).
Also, star and rectangular marks denote the EE performance of the SE maximization scheme $\eta_{SE}$ obtained by the saturation power $P^*$ and $P_{prop}$, respectively.
It is noted that the saturation power for the lower and upper bounds of the EE is the same as the values calculated by (\ref{P_LB}) and (\ref{P_UB}), respectively.
\begin{figure}[t]\centering
\includegraphics[width=5in]{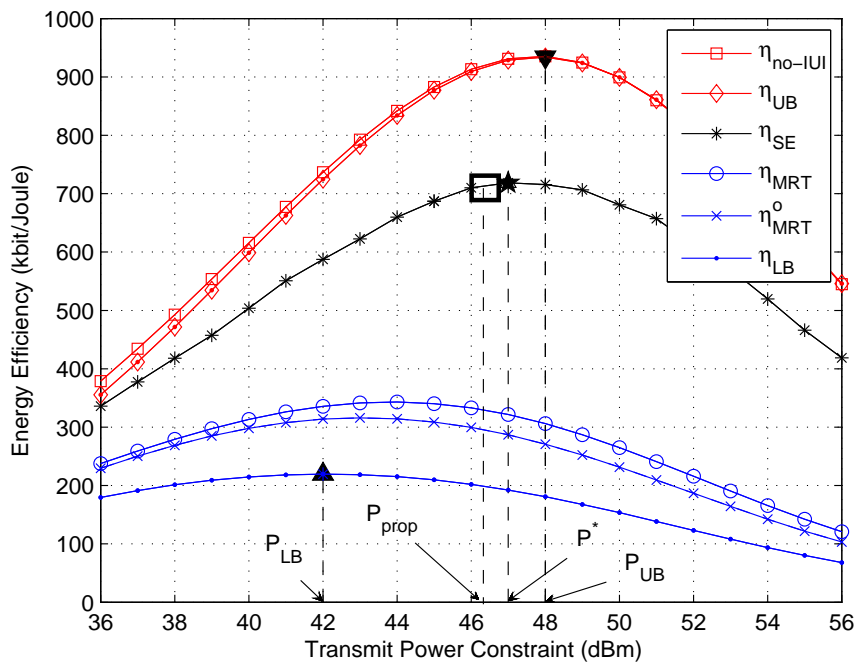}
\caption{Comparison of the saturation power with $P_c' = 40$ dBm and $P_o' = 50$ dBm}
\label{fig3}
\end{figure}
Moreover, $P_{prop}$ is quite well matched with the true saturation power $P^*$.
This demonstrates that our approach of determining the saturation power generates an accurate estimate of the actual saturation power.
When computing $P_{prop}$, we find that $\beta = 1.3$ achieves the maximum EE performance through numerical simulations.
Note that the optimal $\beta$ may change when system parameters vary.
Nevertheless, it can be observed that the proposed saturation power $P_{prop}$ using the fixed $\beta$ yields performance nearly identical to that of $P^*$ for various conditions.

Figure 4 exhibits the comparison of the saturation power for $P_c' = 40$ dBm and $P_o' = 50$ dBm.
Again in this figure, the saturation power derived by (\ref{P_LB}) and (\ref{P_UB}) match well with high accuracy.
In this case, $P^*$ is slightly larger than $P_{prop}$.
Despite the gap between the saturation power, the EE performance corresponding to $P_{prop}$ is very close to the maximum EE in \cite{He:13}.
Moreover, the average EE performance $\eta_{UB}$ and $\eta_{MRT}^\circ$ obtained from the large system analysis are quite close to that of $\eta_{no-IUI}$ and $\eta_{MRT}$ for the finite system case, respectively.
Therefore, we can conclude that even for a system with finite dimension, the analysis of the EE performance with the large system limit provides an accurate approximation.
\begin{figure}[t]\centering
\includegraphics[width=5in]{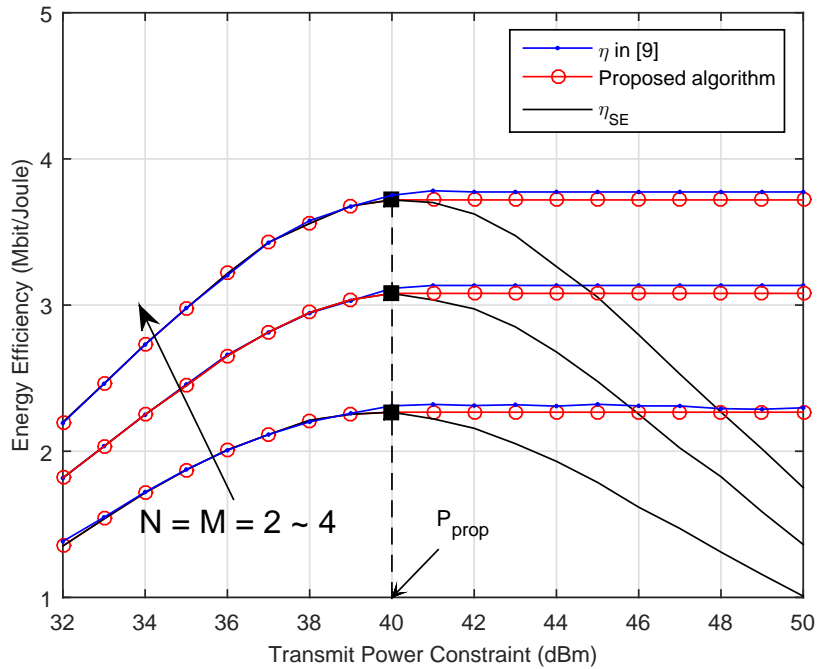}
\caption{EE performance of the proposed algorithm with $P_c' = 30$ dBm and $P_o' = 40$ dBm}
\label{fig4}
\end{figure}

Next, we validate the EE performance of the proposed scheme based on the derived saturation power in Figure 5 with $N=M$.
In this figure, it is observed that the EE performance becomes larger when $M$ and $N$ are increased from 2 to 4.
Note that compared to the EE maximization algorithm in \cite{He:13}, almost the same EE performance is achieved by the proposed method with $P_{prop}$ which utilizes the SE maximization scheme with much reduced complexity.
Also, in Figure 6, we demonstrate the EE performance for $P_c' = 40$ dBm and $P_o' = 50$ dBm.
\begin{figure}[t]\centering
\includegraphics[width=5in]{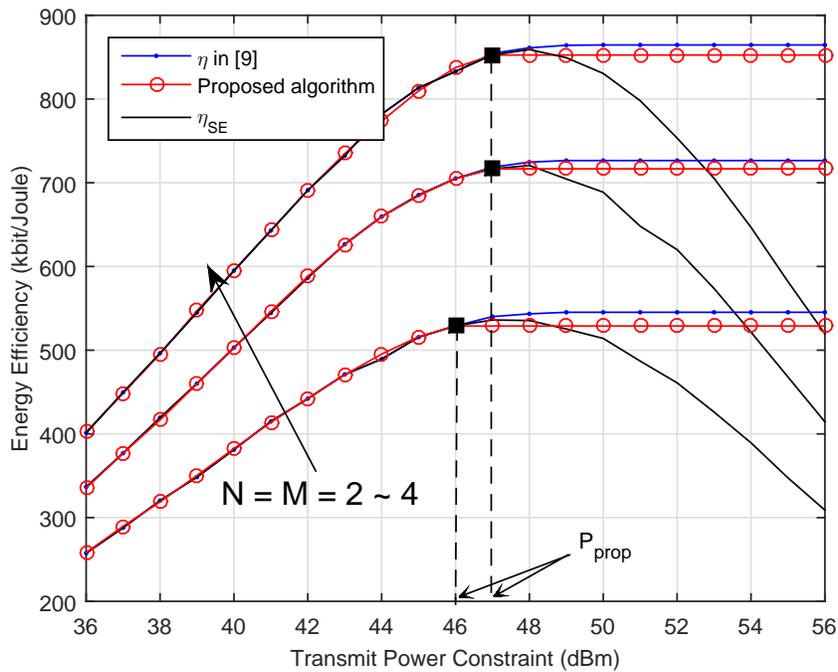}
\caption{EE performance of the proposed algorithm with $P_c' = 40$ dBm and $P_o' = 50$ dBm}
\label{fig5}
\end{figure}
It is remarkable that the proposed scheme with $P_{prop}$ produces the EE performance quite close to the optimal EE solution in \cite{He:13} for different configurations.
Furthermore, the derived saturation power gives insight for the BS power designs in terms of the EE.

\begin{figure}[t]\centering
\includegraphics[width=5in]{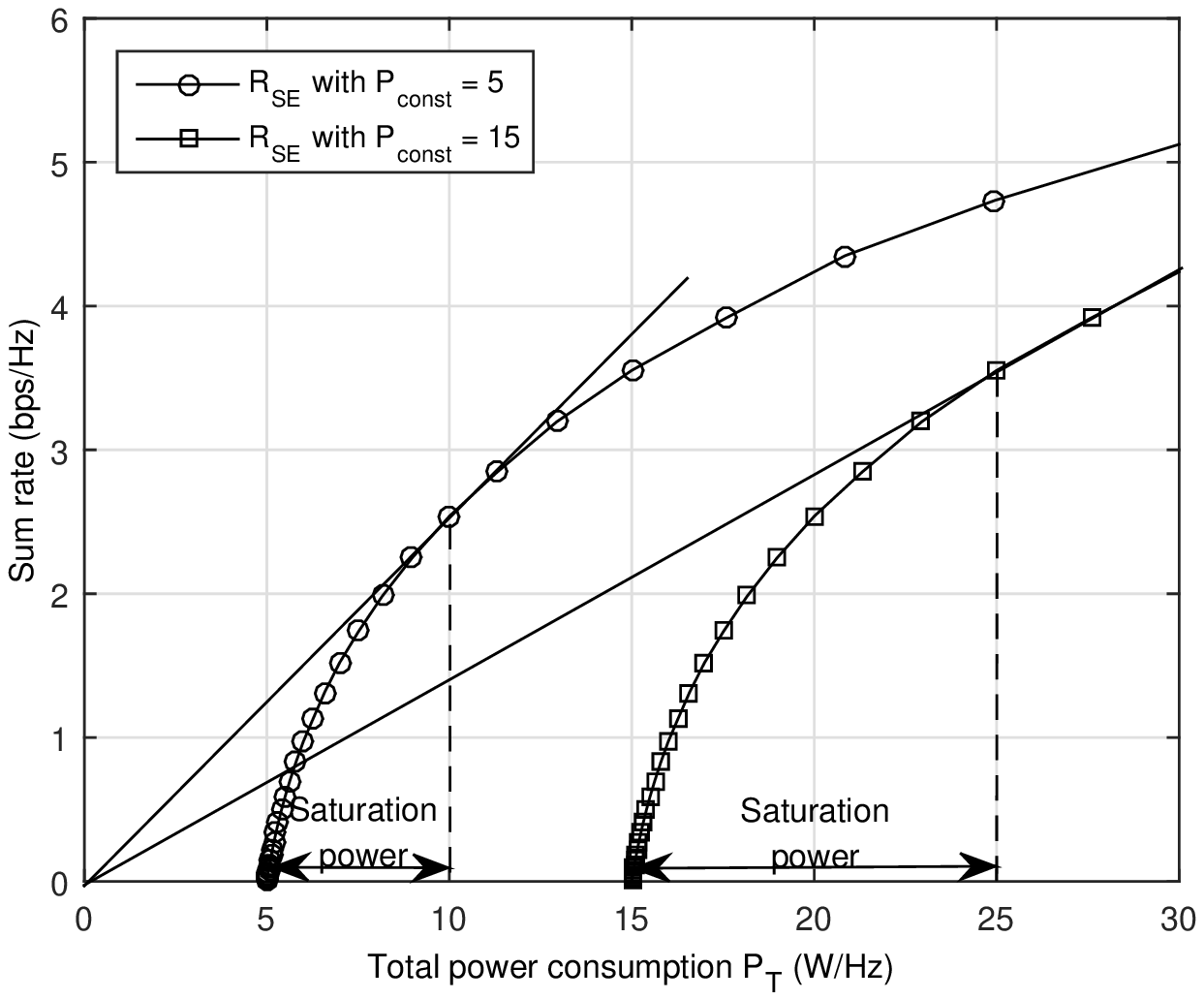}
\caption{The impact of the EE performance and the saturation power with respect to constant power consumption $P_{const}$}
\label{fig6}
\end{figure}
Finally, we exhibit the effect of the EE performance with respect to the constant power consumption term $P_{const}$.
Comparing Figures 5 and 6, when $P_{const}$ increases, the saturation power for achieving the maximum EE also becomes large while the performance of the EE is decreased.
In Figure 7, this phenomenon is illustrated by the trade-off curve of the sum rate and the total consumption power.
In the plot, the curves with circular and rectangular marks denote the sum rate for the SE maximization scheme with $P_{const} = 5$ and 15, respectively.
We can see that for a larger $P_{const}$, the trade-off curve is shifted to the right.
Then, the optimal slope of the tangent which accounts for the performance of the EE becomes small.
On the contrary, the required saturation power is increased to achieve the optimal slope of the tangent.
From these results, we confirm that reducing the amount of $P_{const}$ has a main impact on improving the performance of the EE and saving the transmit power consumption.

\section{Conclusions} \label{sec:conclusion}
In this paper, we have proposed a simple scheme to solve the EE maximization problem for MU-MISO channels.
Leveraging the relationship between EE and SE, the EE is maximized by only utilizing the SE maximization scheme based on the saturation power.
From large system analysis, we have determined the saturation power corresponding to the maximal EE in closed form by exploiting the property between lower and upper bounds of the EE.
This asymptotic result provides insight into the saturation power of the EE for various system configurations.
As a result, the proposed EE scheme makes it possible to provide solutions for the EE maximization efficiently.
It is noted that a performance loss of the proposed scheme is quite small compared to the optimal EE maximization scheme in \cite{He:13}, and the computational complexity of the proposed scheme is significantly reduced.
Also, the simulations demonstrate that the asymptotic results are well matched even for the finite system case.
\bibliographystyle{ieeetr}
\bibliography{AZREF}
\end{document}